\documentclass[reprint,groupedaddress]{revtex4-1}

\usepackage{graphicx}
\usepackage{dcolumn}
\usepackage{bm}%
\usepackage{physics}
\usepackage{tabularx}
\usepackage{xcolor}
\usepackage[colorlinks=true, allcolors=blue]{hyperref}

\begin{document}

\title{Valley transitions during spin shuttling in Si/SiGe heterostructures}


\author{Jonas R. F. Lima}
\affiliation{Department of Physics and IQST, University of Konstanz, 78457 Konstanz, Germany }

\author{Guido Burkard}
\affiliation{Department of Physics and IQST, University of Konstanz, 78457 Konstanz, Germany }


\date{\today}

\begin{abstract}

Valley excitations are considered one of the biggest challenges for a coherent spin shuttling in Si/SiGe heterostructures. In this work we theoretically investigate the valley dynamics during spin shuttling in a Si/SiGe heterostructure by considering a two-level system given by the two low-lying valley states of the mobile quantum dot. We assume that the electron is initially in the valley ground state and analyze the evolution of this state during shuttling. We find that, depending on the shuttling velocity, the system can be in three distinct regimes. In one of these regimes, the quantum information is preserved even in the presence of the valley excitations, allowing for coherent spin shuttling independently of the shuttling distance. This regime is obtained for a shuttling velocity of tens of m/s, a desired regime where dephasing mechanisms, such as spin-valley hotspots and spatially varying spin splitting, can be suppressed. This result is in accordance with recent experimental work that obtained a high spin shuttling fidelity after a fast and long shuttling. We also propose strategies based on the Partial Landau-Zener model to enhance the valley shuttling fidelity, which was found to be much more efficient than previously proposed strategies based on the standard Landau-Zener model. 

\end{abstract}

\pacs{}

\maketitle

\section{Introduction}

Highly isotopically purified Si/SiGe quantum dots are excellent hosts for semiconductor spin qubits \cite{RevModPhys.85.961,RevModPhys.95.025003}. The weak spin-orbit coupling and the zero-spin isotopes enable long relaxation \cite{Morello,Yang,PhysRevApplied.11.044063} and dephasing \cite{PhysRevB.83.165301,Tyryshkin,doi:10.1126/science.1217635} times and high fidelities for single and two-qubit gates have already been experimentally demonstrated \cite{Veldhorst,Yoneda,doi:10.1126/science.aao5965,Watson,Huang,Xue,Noiri,doi:10.1126/sciadv.abn5130}. However, scalable architectures require a mechanism to connect distant spins to lower resource overheads for quantum error correction, and, for semiconductor qubits, due to the wiring bottleneck limitation of the devices. The intrinsic exchange coupling between spin qubits is strong but short-range. A promising solution that has been experimentally implemented by several research groups is the use of mobile spin qubits \cite{fujita2017,mills2019,yoneda2021,noiri2022,seidler2022,PRXQuantum.4.030303}, where the spins can be shuttled for long distances on the device. 

Most of the decoherence mechanisms during spin shuttling, such as spin-valley hotspots and spatially varying spin splitting, are expected to be sufficiently suppressed to allow fault-tolerant quantum computing for a shuttling velocity $v \gtrsim 10$~m/s \cite{PRXQuantum.4.020305}, which has already been surpassed in some of the recent experiments \cite{delftexp,volmer2025}. In this way, one of the main remaining challenge for long-range and fast shuttling operations are valley excitations. The valley degree of freedom emerges because of the presence of six degenerate minima in the conduction band of silicon \cite{RevModPhys.85.961}. This valley degeneracy is partially lifted by strain in a Si/SiGe heterostrucutre, leaving a two-fold degeneracy of valleys along the growth direction, which is typically lifted weakly (typically by tens to hundreds of $\mu$eV) by interface and confinement effects. In Si/SiGe, the valley splitting between the two low-lying valley states is determined by the random distribution of Ge atoms at the Si/SiGe interface \cite{PaqueletWuetz2022}, which means that it changes randomly as the quantum dot moves through the device during shuttling. Therefore, uncontrolled transitions between the two quasi-degenerate valley states may occur. Due to the different $g$-factor of the valley states \cite{ferdous2018,PhysRevB.98.245424}, a valley excitation causes dephasing when operated in the presence of a magnetic field. 

In order to suppress the valley-induced dephasing, some strategies have been proposed recently to mitigate valley excitations during spin shuttling, which also include strategies to increase the valley splitting \cite{losert2023practical,woods2024,thayil2025}. It was already proposed, e.g., to elongate the quantum dot along the shuttling direction \cite{losert2024}, to avoid the crossing of valley splitting minima by allowing the electrons to detour around them \cite{nemeth2025}, or to reduce the shuttling velocity in a small segment of the shuttling path \cite{oda2024}.  These strategies are based explicitly or implicitly on the standard Landau-Zener model that predicts a high probability of a valley excitation after crossing a valley splitting minimum. However, it was found recently that the valley excitations are better described by a generalization of the LZ model, called the Partial Landau-Zener model \cite{lima2025}. The reason for this is that the valley phase plays the most important role in the valley excitations, and strategies based on this model are still missing in the literature. Additionally, the evolution of the valley excitations during shuttling, which can elucidate how these excitations affect quantum information stored on the spin, has not been considered yet.

To understand the valley dynamics during shuttling in SiGe heterostructures, we consider here a two-level system given by the two low-lying valley states of the SiGe quantum dot and investigate the evolution of the valley excitations during shuttling. We first calculate the valley splitting and its phase as a function of the quantum dot location using the model proposed in Refs.~\cite{Lima_2023,lima2024}. Then, we solve the model and define the valley shuttling fidelity. Instead of the expected adiabatic and non-adiabatic regimes for slow and fast shuttling, respectively, we obtain three distinct regimes for the valley dynamics as a function of the shuttling velocity. One of these regimes is ideal for spin shuttling, since the valley excitations in this case do not destroy the quantum information, irrespective of the shuttling distance. This regime is obtained for a shuttling velocity of tens of m/s, which is a desired range for the shuttling velocity. Our findings are in accordance with recent experimental results \cite{delftexp,volmer2025} that obtained a high spin shuttling fidelity, even after a fast and long spin shuttling, revealing that the predicted dephasing induced by the valley excitations is not taking place. In addition, we also propose three new and highly efficient strategies based on the partial Landau-Zener model to enhance the fidelity of shuttling in the presence of a valley quasi-degeneracy.

This paper is organized as follows. In Sec.~\ref{sec:model} we calculate the valley splitting and its phase as a function of the location of the quantum dot for typical Si/SiGe devices with a disordered SiGe barrier region. We solve the quantum-dot two-valley system and define the valley shuttling fidelity. In Sec.~\ref{sec:fidelity} we show the results for the valley shuttling fidelity as a function of the shuttling velocity and describe the three regimes for the valley dynamics during shuttling. In Sec.~\ref{sec:strategies} we propose three strategies based on the partial Landau-Zener model to enhance the valley shuttling fidelity and compare them with previously proposed strategies based on the standard Landau-Zener model. The paper is concluded and summarized in Sec.~\ref{sec:conclusions}.

\section{Theoretical Model}
\label{sec:model}

\subsection{Valley splitting}

Within effective mass theory, the two low-lying valley states of the silicon quantum dot can be written as \cite{PhysRevB.75.115318}
\begin{equation}
|\pm z\rangle = \Psi_{xyz}e^{\pm ik_0z}u_{\pm z}(\textbf{r}),
\end{equation}
where $ \Psi_{xyz}$ is the envelope function, $u_{\pm z}(\textbf{r})$ is the periodic part of the Bloch wavefunctions, $k_0 = 0.82(2\pi /a_0)$ is the Bloch wavenumber at the conduction band minima of silicon, and $a_0 = 0.543$~nm is the length of the Si cubic unit cell. 

Using the model for the calculation of the valley splitting proposed in Refs. \cite{Lima_2023,lima2024}, the Hamiltonian describing the envelope function of the system is given by
\begin{eqnarray}
H=\frac{p_x^2}{2m_t} + \frac{1}{2} m_t \omega_x^2 x^2 + \frac{p_y^2}{2m_t} + \frac{1}{2} m_t \omega_y^2 y^2 \nonumber \\
+ \frac{p_z^2}{2m_l} -eF_zz + U(x,y,z),
\label{H}
\end{eqnarray}
where $m_t = 0.19 \; m_e$ and $m_l = 0.98 \; m_e$ are the transverse and longitudinal effective masses and $\omega_x = 2\hbar/m_tx_0^2$ and $\omega_y = 2\hbar/m_ty_0^2$ are the confinement frequencies along $\hat{x}$ and $\hat{y}$ directions, with $x_0$ and $y_0$ being the size (radius or semi-axis) of the quantum dot along $\hat{x}$ and $\hat{y}$.  We consider in all results here a strong electric field of $F_z = 10$~MV/m along the growth (stacking) direction of the heterostructure. The potential $U(x,y,z)$ is given by a sum of delta functions at the locations of each of the Ge atoms in the barrier region,
\begin{equation}
U(x,y,z)=\lambda \sum_i \delta(x-x_i) \delta(y-y_i) \delta(z-z_i),
\label{U}
\end{equation}
where $\lambda = 10$~meV$\cdot$nm$^3$ is a parameter of the model fixed by the known band offset of $150$~meV between Si and Si$_{0.7}$Ge$_{0.3}$, and $i$ labels the Ge atoms.

Considering a uniform distribution of Ge atoms in the $x$ and $y$ directions, we can write
\begin{eqnarray}
U(x,y,z)= \bar{U}(z)+\delta U(x,y,z),
\end{eqnarray}
where we can treat $\delta U$ as a small perturbation. Therefore, the unperturbed Hamiltonian becomes separable. In the $z$ direction, we distribute the Ge atoms following a probability distribution function given by a hyperbolic tangent function, which means that
\begin{eqnarray}
\bar{U}(z)= \frac{U_0}{2}[\tanh ((-d_w-z)/L_b)+1] \nonumber \\
+\frac{U_0}{2}[\tanh (z/L_t)+1],
\label{ubar}
\end{eqnarray}
where $U_0 = 150$~meV and $L_b$ and $L_t$ control the widths of the bottom and top interfaces, respectively. Since the electric field pushes the envelope function towards the upper SiGe barrier, the bottom interface will have a negligible contribution to the valley splitting. So, from now on, we write $L_t = L$, where the width of the interface is given by $w = 2L$.

The intervalley coupling is given by
\begin{eqnarray}
\Delta &=& \langle +z |U(x,y,z)|-z\rangle. \nonumber \\
&=&C_0\int e^{-2ik_0z}U(x,y,z)|\Psi_{x,y,z}|^2 d^3x,
\end{eqnarray}
where $C_0 = -0.2607$ comes from the periodic parts of the Bloch wavefunctions \cite{PhysRevB.84.155320,PhysRevResearch.2.043180}. 
The valley splitting is then given by
\begin{eqnarray}
E_{\rm VS} = 2 |\Delta|.
\label{Evs}
\end{eqnarray}

\begin{figure}[h]
\includegraphics[width=\linewidth]{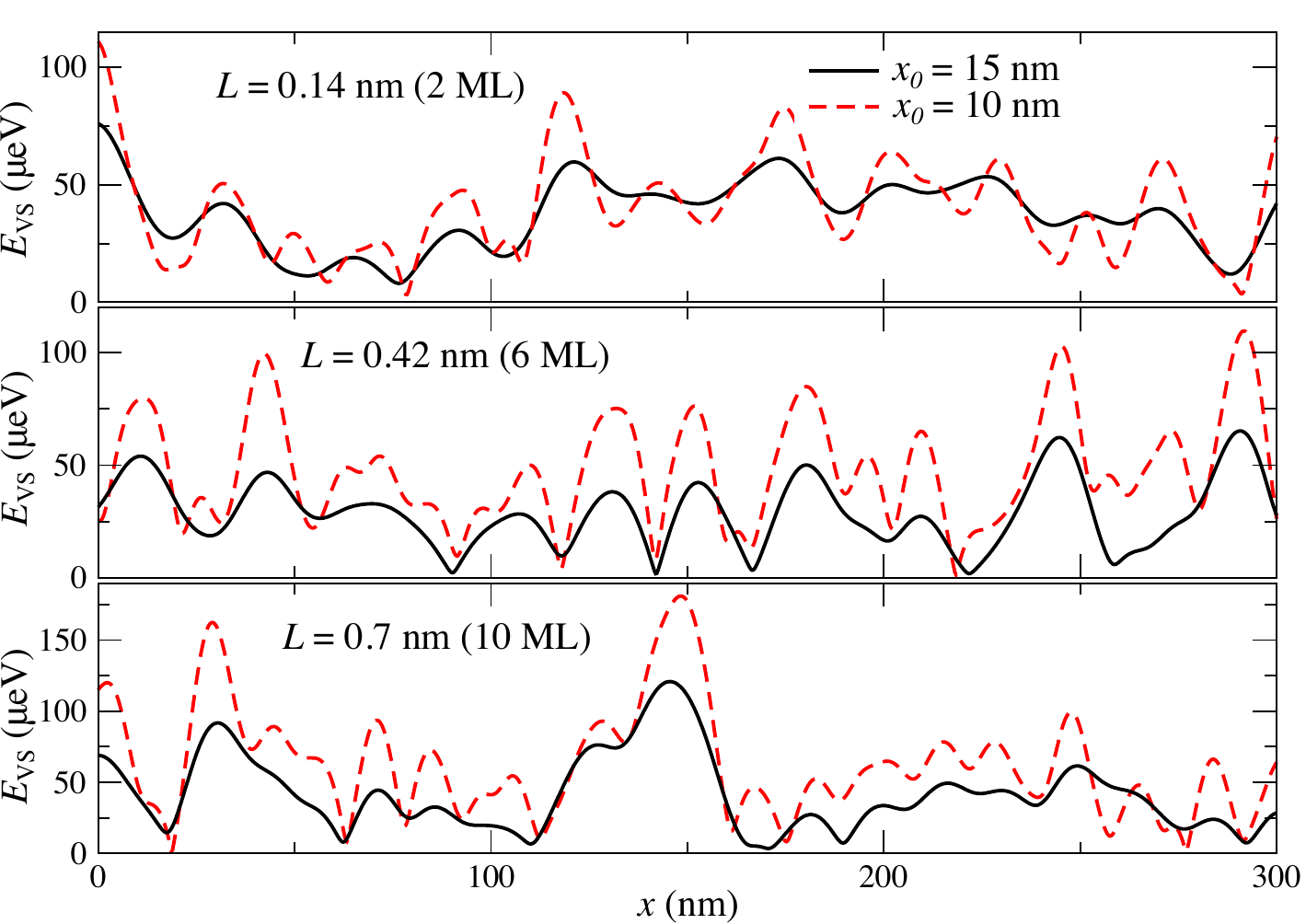}
\caption{\small Valley splitting as a function of the location of the quantum dot for a shuttling distance of 300 nm in the $x$ direction. We consider here two values for the radius of the quantum dot, 15 nm (solid black lines) and 10 nm (dashed red lines), and three values for the interface width, 2 ML (top panel), 6 ML (middle panel) and 10 ML (bottom panel).} \label{vs}
\end{figure}

We consider here three values for the interface width (2, 6 and 10 monolayers (ML), where one ML has a width of $\approx 0.14$~nm) and a quantum dot radius $x_0$ of 15~nm. The valley splitting for all these cases as a function of the location of the quantum dot for a shuttling path of 300 nm in the $x$ direction can be seen in Fig.~\ref{vs}. In all cases, the valley splitting shows random oscillations during the shuttling. The amplitude of these oscillations decreases as we decrease (increase) the interface width (the quantum dot radius), which is in agreement with the results obtained in Ref.~\cite{lima2024}.  Such reduction in the amplitude of the oscillations also leads to a smoother passage through the valley splitting minima, which may reduce the probability of the valley excitation. 

Based on the standard LZ model, it is believed that the use of strategies to enhance valley splitting would reduce the probability of valley transitions. This can be done, e.g., by reducing the size of the quantum dot. For this reason, we also consider here the case with $x_0 = 10$~nm. As can be seen in the dashed red lines in Fig.~\ref{vs}, the reduction of the dot size increases the valley splitting in many regions, but it does not remove the valley splitting minima. This happens because a reduction of the dot radius increases the probability of the appearance of high valley splittings, but it does not completely remove the probability of small values of valley splitting. 
\begin{figure}[b]
\includegraphics[width=\linewidth]{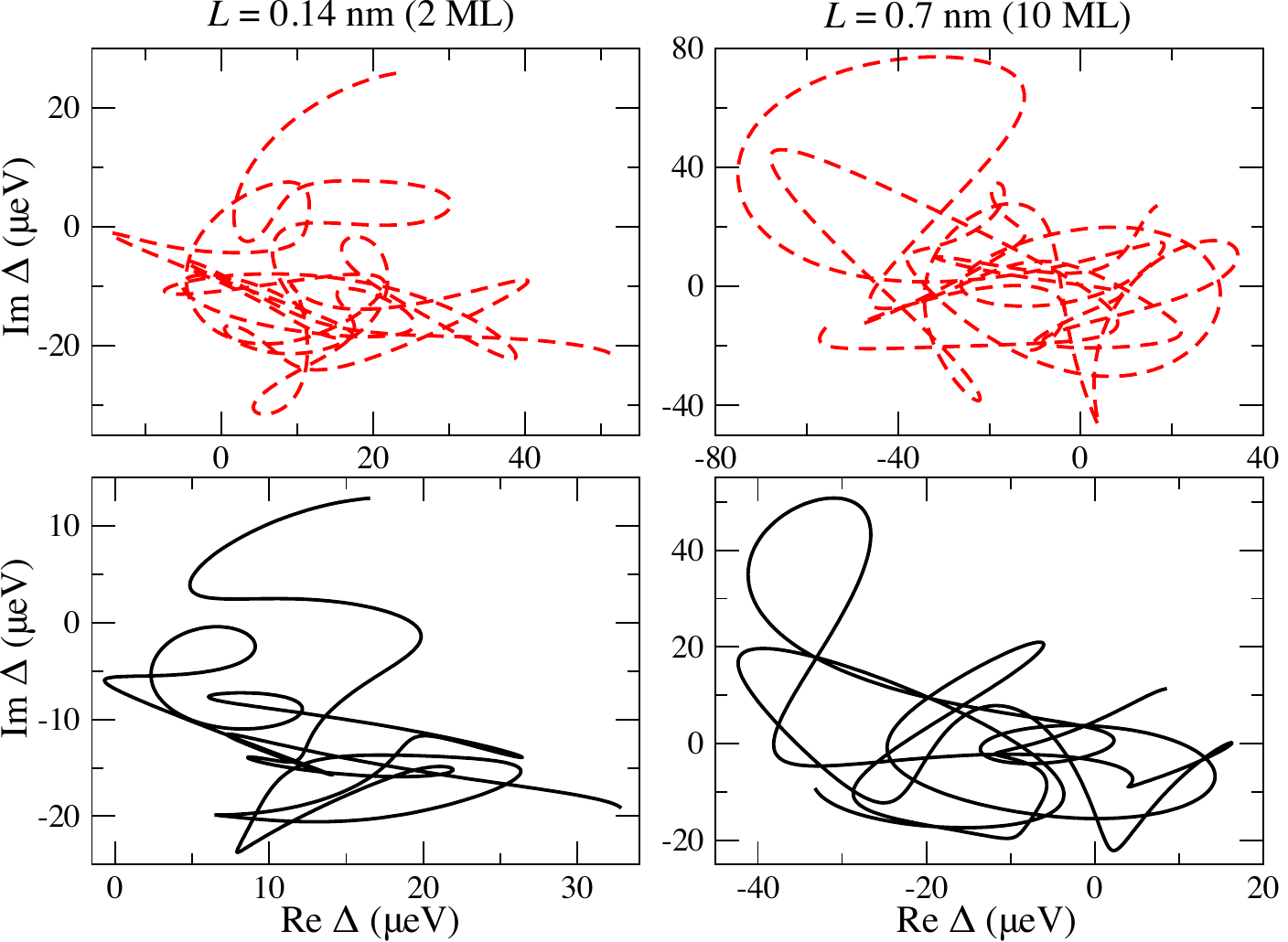}
\caption{\small The change in the complex plane of the intervalley coupling $\Delta$ as we move the quantum dot. We consider here two values for the interface width and two possible radii of the of the quantum dot. The colors and line styles are the same as in Fig.~\ref{vs}.} \label{vsphase}
\end{figure}

Besides the valley splitting, the probability of a valley excitation will also be affected by the valley phase, which is what we call the argument of the complex intervalley coupling $\Delta$. Figure~\ref{vsphase} reveals how the intervalley coupling changes in the complex plane as the quantum dot moves along the x axis. In addition to the large amplitude of the oscillations of the valley splitting for the cases with small $x_0$ and large interface width, we also find a greater change in the valley phase for these cases. This is manifested by the large number of times that the intervalley coupling moves to different quadrants in the complex plane. Such great valley phase changes may favor valley excitations, as we discuss in more detail below.

\subsection{Valley transitions and the Landau-Zener model}

Noticing how the valley splitting changes in Fig. \ref{vs}, we are led to describe each valley splitting minimum as an avoided crossing point of the Landau-Zener problem. In this way, we could use the Landau-Zener formula to predict the probability of a valley excitation at each of these points. However, the Landau-Zener formula is not valid here \cite{lima2025}. 

\begin{figure}[h]
\includegraphics[width=\linewidth]{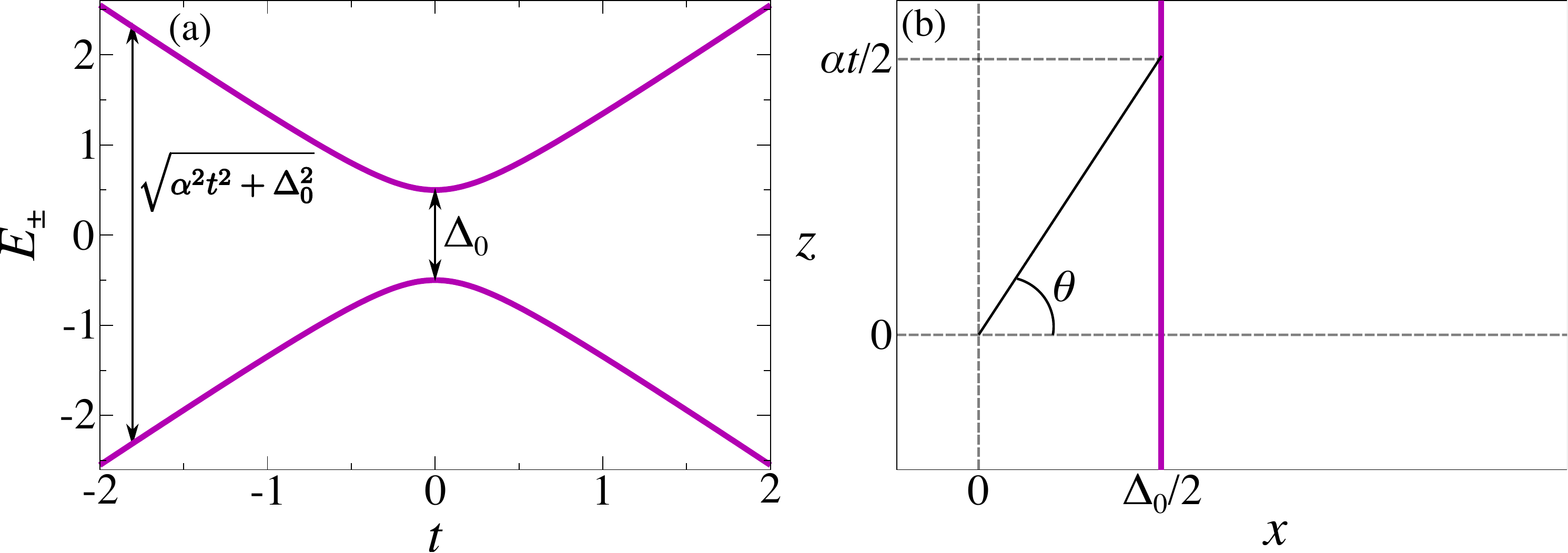}
\caption{\small A generic two level system described by the standard Landau-Zener model. (a) The energy levels around the avoided crossing point at $t=0$ with an energy gap of $\Delta_0$. (b) The Hamiltonian curve, that is a straight line in this case.} \label{SLZ}
\end{figure}

The most general Hamiltonian for a two level system can be written as
\begin{equation}
H_{2L} = -\left( \begin{array}{cc}
z & x-iy \\
x+iy & -z
\end{array}\right).
\label{hgeneral}
\end{equation}
For the standard Landau-Zener problem we have $y=0$, $x=\Delta_0/2$ and $z = \alpha t/2$, where $\Delta_0$ is the constant energy gap between the two-level system at the avoided crossing point and $\alpha$ is the driving velocity, which can be seen in Fig. \ref{SLZ} (a). By considering that the system is initialized in the infinite past $t \rightarrow -\infty$ in the ground state, and evolves with a constant velocity to $t \rightarrow \infty$ with the avoided crossing point at $t = 0$, the probability of a transition to the excited state is given by the formula
\begin{equation}
P_D = e^{-2\pi \Gamma}
\end{equation}
where
\begin{equation}
\Gamma = \frac{\Delta_0^2}{\hbar |\alpha|}.
\end{equation}

If we consider a driving restricted to a finite range of time, namely, from $-t_a$ to $t_a$, depending on the parameters of the system and the total driving time, the Landau-Zener formula can still be a good approximation for the probability of a transition. However, this approximation is valid when the Hamiltonian curve $(x(t),y(t),z(t))$ is a straight line, which can be seen in Fig \ref{SLZ} (b).  The reason is because the Landau-Zener formula is obtained from the asymptotic expansion of the parabolic cylinder functions, which are the exact solutions of the Hamiltonian (\ref{hgeneral}) for a driving in a straight line Hamiltonian curve.

In our case, the valley Hamiltonian is given by
\begin{eqnarray}
H = \left( \begin{array}{cc}
 0 & \Delta(t)  \\ 
\Delta^*(t) & 0 
\end{array} \right),
\label{hv}
\end{eqnarray}
which means $z=0$ and $x$ and $y$ equal to the real and imaginary parts of $\Delta$, respectively., in the Hamiltonian (\ref{hgeneral}). The main difference to the original Landau-Zener problem is that now both $x$ and $y$ are time dependent. The consequence of this is a Hamiltonian curve different from a straight line, as can be seen in Fig.~\ref{vsphase}. Therefore, the Landau-Zener formula does not apply here.

The valley splitting and the shuttling velocity will still play important roles in valley excitations. However, the main contribution comes from the time evolution of  $\theta(t)$.  In the LZ model, $\theta = \arctan(\alpha t/\Delta_0)$. So, the angular velocity at the anticrossing point is $\dot{\theta}(0)=\alpha/\Delta_0$. For this reason, the transition probability depends only on the driving velocity and the energy gap. In our case, we have that
\begin{equation}
\label{eq:valleyphase}
\theta = \arctan\frac{\Im \Delta}{\Re \Delta},
\end{equation}
and the value of $\dot{\theta}$ at the anticrossing points depends now on the Hamiltonian curve in the complex plane. As we will show here, the best strategies to reduce the probability of a valley excitation relies on reducing $\dot{\theta}$.



\subsection{Valley shuttling fidelity}

In order to obtain the transition dynamics between the two valley states, we consider here a two-level system given by the two valley states with a time-dependent Hamiltonian. This approach does not take into account the contribution of the spin on the valley dynamics. However, recent works obtained that the spin contribution is weak compared to the intervalley coupling, which means that our results here describe accurately the valley evolution. 

For a stationary quantum dot, the Hamiltonian for the two valley states is given by Eq.~(\ref{hv}). The two pure valley states are given by
\begin{eqnarray}
\alpha_{\pm} = \frac{1}{\sqrt{2}}\left( 
\begin{array}{c}
e^{i\theta/2} \\
\pm e^{-i\theta/2}
\end{array}\right),
\end{eqnarray}
which will be instantaneous valley states. The valley phase $\theta$ is given in Eq.~(\ref{eq:valleyphase}). Diagonalizing Eq.~(\ref{hv}), we obtain the energy eigenstates as
\begin{equation}
E_{\pm} = \pm |\Delta|,
\end{equation}
which gives rise to the valley splitting in Eq.~(\ref{Evs}).

For a moving quantum dot, we solve the time-dependent Schr\"odinger equation
$i\hbar \partial_t \psi(t) = H(t)\psi(t)$,
which for a constant shuttling velocity $v$ with $x=vt$ takes the form
\begin{equation}
i\hbar v \partial_x \psi(x) = H(x)\psi(x).
\label{hx}
\end{equation} 

We solve Eq.~(\ref{hx}) numerically considering that initially the electron is in the valley ground state $\alpha_-$ and write
\begin{equation}
\ket{\psi} = a\ket{\alpha_-}+b\ket{\alpha_+}.
\end{equation}
The valley shuttling fidelity is then given by
\begin{equation}
F_V = |\bra{\alpha_-}\ket{\psi}|^2 = |a|^2.
\end{equation}

\section{Valley shuttling fidelity $F_V$}
\label{sec:fidelity}

\begin{figure}[h]
\includegraphics[width=\linewidth]{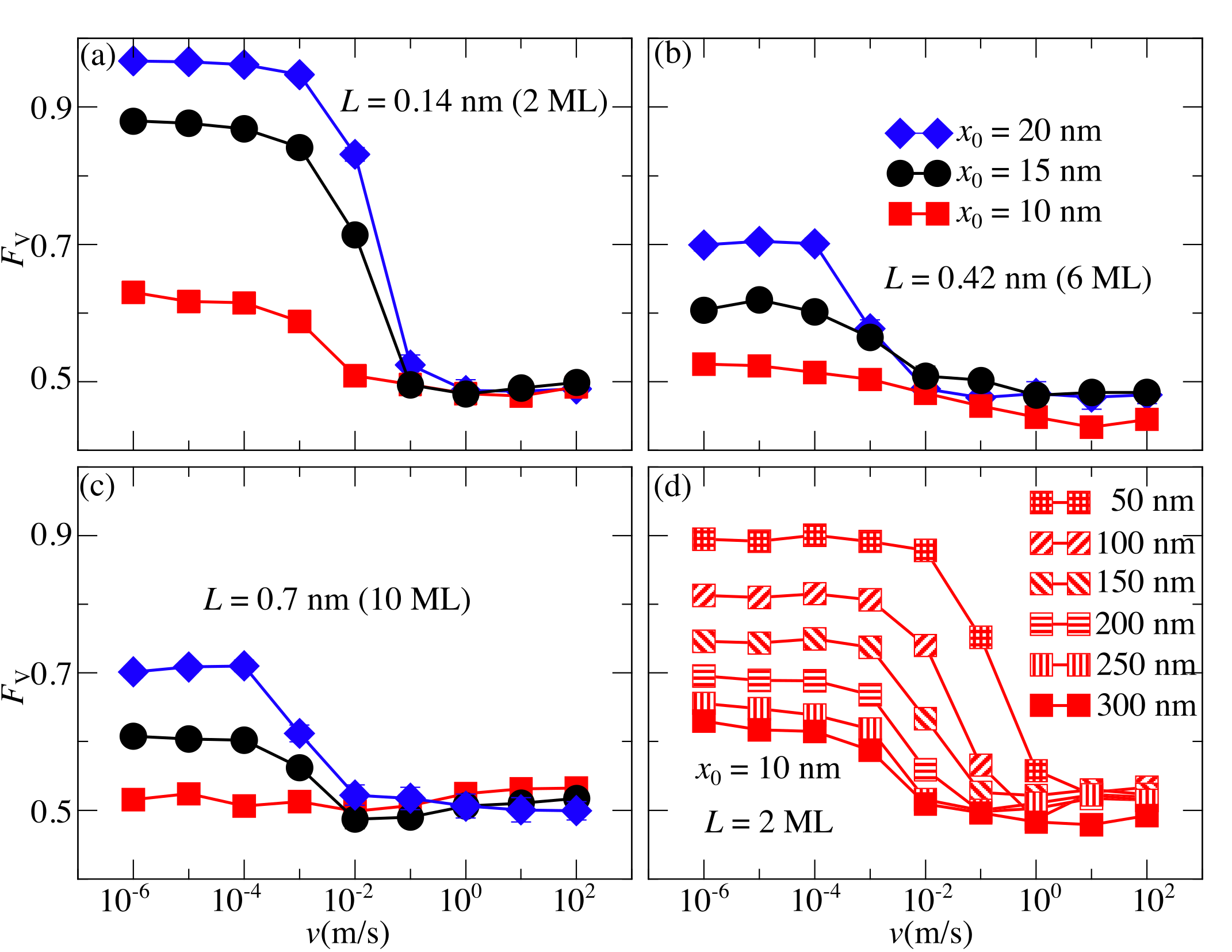}
\caption{\small The average shuttling valley fidelity as a function of the shuttling velocity for all cases considered here . In (a), (b) and (c) we have the devices with interface with of 2, 6 and 10 ML, respectively, and a shuttling distance of 300 nm. In (d) we compare the average shuttling valley fidelity for different values of the shuttling distance for a device with a quantum dot radius $x_0$ of 10 nm and an interface width of 2 ML.} \label{Fav}
\end{figure}

We first calculate the average valley shuttling fidelity as a function of the shuttling velocity, where the average is taken over 1000 realizations, for dot sizes $x_0 = 10, 15, 20$~nm, as seen in Fig.~\ref{Fav}. It seems that there is a transition between two regimes as the shuttling velocity increases, which can be seen more clearly in the case with $x_0 = 20$~nm. For the cases considered here, the transition between these two regimes occur at a critical shuttling velocity region $v_c$ between $10^{-2}$ and $10^{-1}$~m/s. Let us now characterize these regimes.

An adiabatic regime for small shuttling velocities and a nonadiabatic regime for a fast electron shuttling would be expected. However, when $v < v_c$, we find that $F_V$ decreases with increasing shuttling distance increases, as shown in Fig.~\ref{Fav}(d). This means that even for slow electron shuttling, there will be a valley leakage that increases with the shuttling distance. We call this the \textit{distance-dependent valley leakage} (DVL)  regime. This is not expected based on the standard LZ model. This valley leakage even at very slow shuttling may be a consequence of the valley phase change. Comparing the different model devices, we can see that an increase of the quantum dot size reduces the valley leakage in this regime. This is due to a reduction of the disorder and the variability of the valley splitting with increasing dot size $x_0$. Consequentially, even though a reduction of $x_0$ increases the probability of having regions in the device with high valley splitting, as can be seen clearly in Fig.~\ref{vs}, this also increases the probability of a valley excitation during shuttling. This means that the use of strategies to enhance the valley splitting that do not remove the probability of small VS should increase the spin shuttling infidelity in this regime. Moreover, a sharper interface reduces the probability of a valley excitation, which is also a consequence of the reduction of the valley splitting variability. The DVL regime is not ideal for spin shuttling, since the valley leakage would induce considerable spin dephasing at long shuttling distances. 

The increase of the variability of the valley splitting reduces the average valley fidelity to 0.5, even in the DVL, which can be seen in Fig.~\ref{vs} for the case $x_0 = 10$~nm. In this case, due to the increase of the valley leakage, the distance of 300~nm considered here is enough to have a complete leakage from the valley ground state to the valley excited state. After this, a leakage backwards occurs between the valley states. So, the valley fidelity can be at any value between 0 and 1. Taking the average over the samples, we find $\bar{F}_{V} = 0.5$.

\begin{figure}[h]
\includegraphics[width=\linewidth]{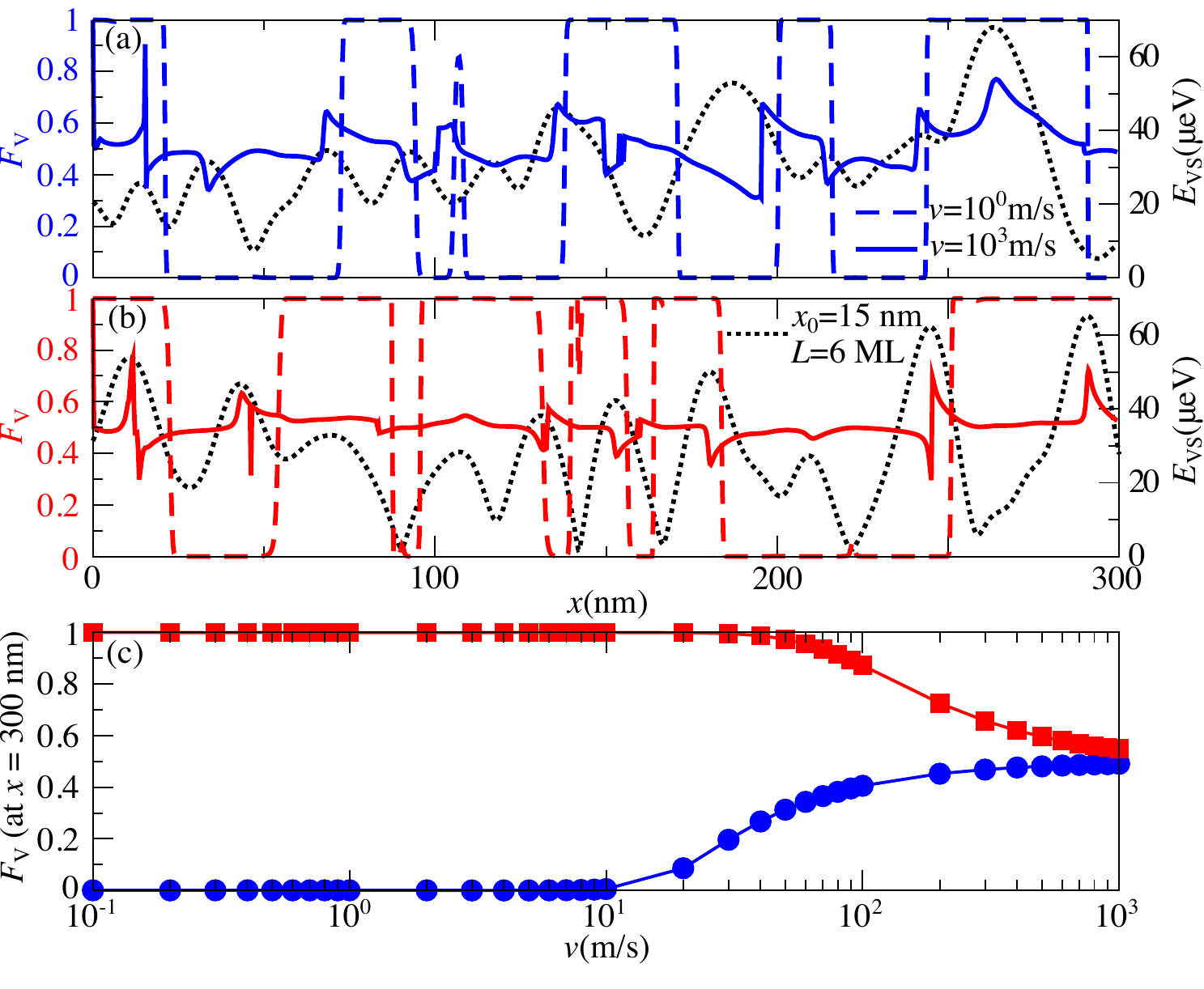}
\caption{\small (a) and (b) The valley shuttling fidelity $F_V$ (blue and red lines) and the valley splitting $E_{\rm VS}$ (black dotted line) as a function as the shuttling distance $x$. Panels (a) and (b)  represent two distinct realizations if the disordered Si/SiGe interface with dot size $x_0=15$~nm and interface width $L=6$~ML. The valley shuttling fidelity is plotted for two different shuttling velocities, $v=10^0$~m/s (dashed line, FVT regime) and $v=10^3$~m/s (solid line, CVS regime). (c) The valley shuttling fidelity at $x = 300$~nm as a function of the shuttling velocity $v$ for the two device realizations considered in panels (a) and (b).} \label{Fv}
\end{figure}

To understand the average shuttling valley fidelity of $0.5$ when $v>v_c$, we plot $F_V$ as a function of the shuttling distance in Fig.~\ref{Fv} for two realizations. We can see that, for $v=1$~m/s there is a sequence of transitions between the two valley states. In some realizations, the system ends up in the valley excited state with $F_V = 0$ (Fig.~\ref{Fv}(a)), and in other realizations, it ends up in the valley ground state with $F_V = 1$ (Fig.~\ref{Fv}(b)). On average, this leads to $\bar{F}_V = 0.5$. We find that this is the case for shuttling velocities between $10^{-1}$~m/s and $10$~m/s, as seen in Fig.~\ref{Fv} $(c)$, where $F_V$ at $x = 300$~nm is plotted as a function of $v$ for the two realizations considered in Fig.~\ref{Fv}(a) and (b). We call this the \textit{full valley transitions} (FVT) regime. For a shuttling velocity higher than $10$~m/s, the electron is in a linear combination of both valleys, still having $\bar{F}_V = 0.5$. We call this the\textit{combined valley states} (CVS) regime. Therefore, we identify a total of three distinct regimes. 

We expect that in the DVL and CVS regimes, there will be spin-valley entanglement, which is a consequence of the electron being in a superposition of the two valley states during shuttling, which in turn leads to dephasing. Hence, the DVL and CVS regimes are not ideal for spin qubit shuttling. However, in the FVT regime, the quantum information is lost only if there is a valley relaxation while the electron is in the valley excited state. As can be seen in Fig.~\ref{Fv}, the electron is in the valley excited state only for a time much shorter than the expected valley relaxation time. Therefore, the only challenge here is to ensure that the electron ends up in the valley ground state after shuttling. If we have this, the valley excitations during shuttling would produce only a deterministic spin rotation and the quantum information is preserved, regardless of the shuttling distance. This is the ideal regime for spin shuttling.

\section{Strategies to enhance $F_V$}
\label{sec:strategies}

After understanding the valley evolution during electron shuttling in SiGe quantum dots, let us now consider strategies to enhance the shuttling valley fidelity. 

\subsection{Controlled slow-down}

\begin{figure}[h]
\includegraphics[width=\linewidth]{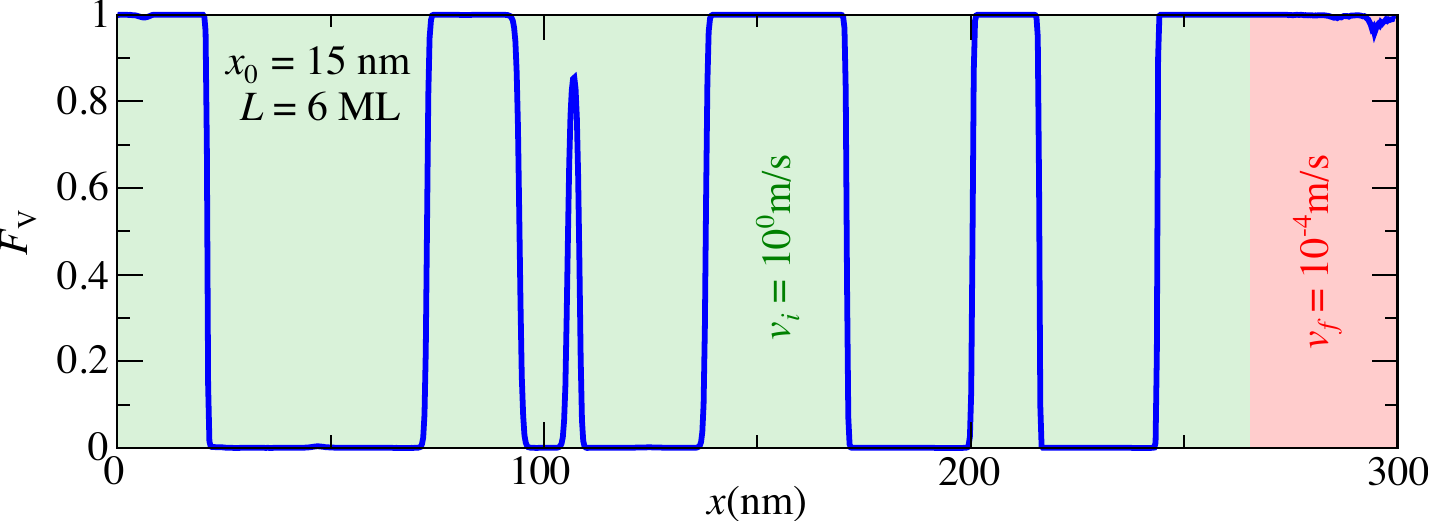}
\caption{\small The valley shuttling fidelity $F_V$ as a function of the shuttling distance $x$ for the device realization considered in Fig.~\ref{Fv}(a). By reducing the shuttling velocity in a small region at the end of the shuttling path (shown in red), it is possible to reach $F_V =1$ instead of $F_V=0$ at the end of the shuttling.} \label{str1}
\end{figure}

We first propose a strategy that can be applied to the FVT regime. As already mentioned, the only challenge here is to ensure that the electron ends up in the valley ground state after shuttling. For a fixed shuttling distance, shuttling in 50\% of the samples already end up in the valley ground state in this regime. For the other samples where shuttling ends up in the valley excited state. In these cases, if one of the valley transitions during shuttling is avoided, then shuttling will result in the valley ground state. This can be achieved by reducing the shuttling velocity in a short region. For instance, in the case showed in Fig.~\ref{Fv}(a) for a shuttling velocity of 10~m/s, it is possible to avoid the last valley transition by reducing the shuttling velocity to $10^{-4}$~m/s during the last phase of the shuttling. After this change of the shuttling schedule, the electron ends up in the valley ground state, as can be seen in Fig.~\ref{str1}. 

A similar strategy was recently proposed in Ref.~\cite{oda2024}. However, this proposal was based on the standard Landau-Zener model, where the shuttling velocity is reduced in a fixed region around a valley splitting minimum. Here, this strategy may work in some samples, but not always. As already discussed, it is only possible to predict the location of the valley transitions if we know both the valley splitting and phase at each point during shuttling. In the cases where the location of the valley transitions is unknown, the best strategy to avoid a single valley transition is to reduce the shuttling velocity in a short region (around 10~nm) at the end of the shuttling and then increase this region little by little until it reaches the location of the last valley transition. 

Since we have a sequence of full transitions between the two valley states in the FVS regime, all samples can end up in the valley ground state after shuttling by simply adjusting the shuttling distance. Therefore, if it is possible to vary the shuttling distance, this controlled slow-down strategy is not necessary. 
 
\subsection{Adapted velocity}

Another strategy to increase the valley shuttling fidelity is to use an adapted velocity. In this case, the idea is to try to avoid valley transitions. We compare two strategies, one based on the standard LZ model and other based on the partial LZ model. 

\subsubsection{Standard Landau-Zener analysis}

As already discussed here, based on the standard LZ model (SLZ), one would expect an increase of the probability of a valley excitation after crossing a valley splitting minimum. The strategy here is to adapt the shuttling velocity in such a way that it increases (decreases) in regions with large (small) valley splitting. To achieve this, we assume that the system is in the DVL regime when $\hbar v/x_0$ is much smaller than the valley splitting. The idea is to keep the system always in this regime, avoiding any valley excitations, and to reduce the distance-dependent valley leakage.  We therefore consider a shuttling schedule given by a position- (and time-) dependent velocity,
\begin{equation}
v(x)=\lambda \frac{x_0 E_{\rm VS}(x)}{\hbar},
\end{equation}
where $\lambda$ controls the relative magnitude of $\hbar v/x_0$ and $E_{\rm VS}$, where $\lambda\ll 1$ for the system to be in the DVL regime. Numerically, we find that this regime is achieved with sufficient precision when $\lambda \lesssim 10^{-5}$. From now on, we will therefore choose $\lambda = 10^{-5}$.

\begin{table}
\caption{Adapted velocity strategy based on the standard Landau-Zener (SLZ) model. In the last row we included the results for a shuttling with a constant velocity of 1~mm/s.}
\begin{tabularx}{0.48\textwidth}{
| >{\centering\arraybackslash\hsize=0.55\hsize}X 
  | >{\centering\arraybackslash\hsize=0.15\hsize}X 
    | >{\centering\arraybackslash\hsize=0.15\hsize}X 
  | >{\centering\arraybackslash\hsize=0.15\hsize}X |}
\hline
interface width $L$ (nm) & $0.14$ & $0.42$ & $0.7$ \\
\hline
shuttling velocity $\bar{v}$ (mm/s) & 6.3 & 4.6 & 6.4 \\
\hline
valley fidelity $\bar{F}_V$ & 0.968 & 0.695 & 0.665 \\
\hline
$\bar{F}_V$  ($v=1$~mm/s) &  0.841  & 0.565 & 0.562 \\
\hline
\end{tabularx}
\label{table1}
\end{table}

In Table~\ref{table1}, we summarize the results for the average valley shuttling fidelity $\bar F_V$ and the average shuttling velocity $\bar{v}$. For comparison, we also included  $\bar{F}_V$ for the case of a constant shuttling velocity of 1~mm/s in the bottom row. In all cases, our adapted velocity strategy outperforms ( in terms of $\bar{F}_V$) the strategy with a constant shuttling velocity, even when the average velocity is higher. In particular, for a heterointerface with $L = 0.14$~nm, it is possible to reach a high valley fidelity of $\bar F_V=0.968$ when moving more than six times faster compared to the constant velocity case for which $\bar F_V=0.841$. This indicates that our strategy can be used to substantially suppress valley leakage during shuttling. Still, it seems only possible to achieve high valley fidelities with an average shuttling velocity much smaller than what is desired in practice.
 
\subsubsection{Partial Landau-Zener analysis}

Let us now consider an adapted velocity strategy based on the partial LZ model which takes into account that the change of the valley splitting phase $\theta$ during shuttling plays an important role in the valley transitions. The adaptive shuttling strategy is to increase (decrease) the shuttling velocity in regions with small (high) angular velocity of the valley splitting phase. 
We write
\begin{equation}
v(x) = \dot{\theta} \left( \frac{d\theta(x)}{dx}  \right)^{-1},
\end{equation}
where $\dot{\theta}=d\theta/dt$. We modulate the shuttling velocity with $\theta(x)$, keeping $\dot{\theta}$ always constant. We found that the DVL regime is achieved when $\dot{\theta} \leq 10^8$~rad/s and consider $\dot{\theta} = 10^8$~rad/s in all results.

In Table~\ref{table2} we summarize our results for $\bar{v}$ and $\bar{F}_V$. One can see that now it is possible to achieve a high valley fidelity in all devices considered here with a much faster average shuttling velocity. For instance, with $L=0.14$~nm, we find a higher valley fidelity, even increasing the shuttling velocity by four orders of magnitude compared to the strategy based on the standard LZ model. This confirms the predictions of the partial Landau-Zener (PLZ) model and also points to a new direction in the search for strategies to suppress spin dephasing during shuttling due to valley excitation.  

\begin{table}
\caption{Adapted velocity strategy based on the partial Landau-Zener (PLZ) model.}
\begin{tabularx}{0.48\textwidth}{
| >{\centering\arraybackslash\hsize=0.55\hsize}X 
  | >{\centering\arraybackslash\hsize=0.15\hsize}X 
    | >{\centering\arraybackslash\hsize=0.15\hsize}X 
  | >{\centering\arraybackslash\hsize=0.15\hsize}X |}
\hline
interface width $L$ (nm) & 0.14 &  0.42 & 0.7 \\
\hline
shuttling velocity $\bar{v}$ (m/s) & 42 & 6.9 & 7.2 \\
\hline
valley fidelity $\bar{F}_V$  & 0.987 & 0.99 & 0.994 \\
\hline
\end{tabularx}
\label{table2}
\end{table}

\subsection{Lateral avoidance}

We also consider another strategy to enhance $F_V$, that involves an additional lateral motion of the quantum dot, whose trajectory then deviates from a straight line. This strategy, which we term lateral avoidance, was considered \cite{nemeth2025} based of the SLZ model, and is based on the avoidance of any valley splitting minima. Here, we compare this SLZ-based strategy with a new strategy based on the PLZ model.

\begin{figure*}
\includegraphics[width=\linewidth]{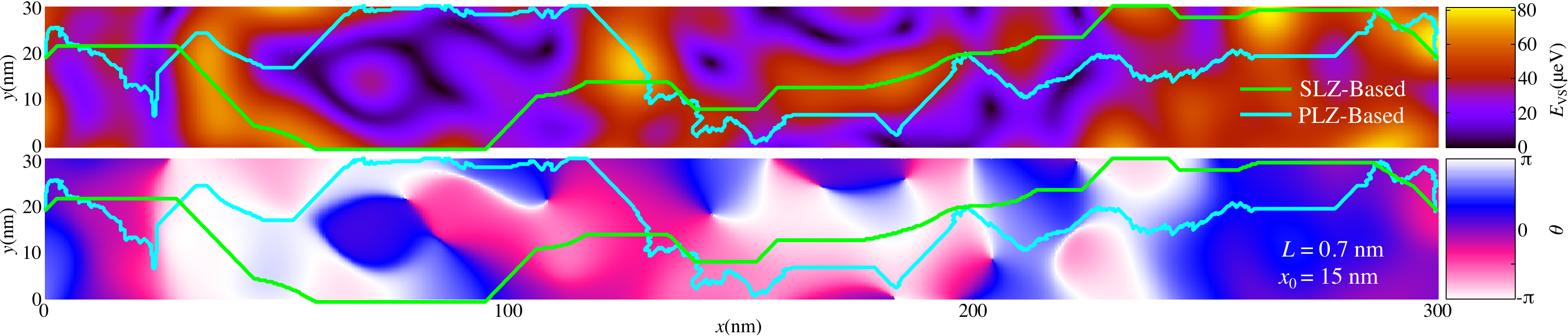}
\caption{\small (Top panel) The valley splitting $E_{\rm VS}$ and (bottom panel) the valley phase $\theta$ (color bar) as a function of the location of the quantum dot $(x,y)$ in the $xy$ plane. We consider a device with an interface width of $L=0.7$~nm and a quantum dot radius of $x_0=15$~nm. We also plot the optimized shuttling paths obtained by the $A^*$ algorithm based on the standard Landau Zener (SLZ) model (green line) and the partial Landau Zener (PLZ) model (cyan line).} \label{2d}
\end{figure*}

In Fig.~\ref{2d}, we show the valley splitting and the valley phase in the $xy$ plane for a shuttling path of 300~nm in the $x$ direction with the possibility of moving in a range of 30~nm in the lateral ($y$) direction. We consider a device with an interface width of $L = 0.7$~nm, and a dot size of $x_0 = 15$~nm. Although the arctangent function that appears in Eq.~(\ref{eq:valleyphase}) is defined in the range ($-\pi/2,\pi/2$), the intervalley coupling moves in the four quadrants of the complex plane during shuttling. For this reason, the valley phase $\theta$ is defined here in the interval ($-\pi,\pi$).

To optimize the shuttling path (compared to a simple straight line), we use the $A^*$ algorithm, which finds the shortest path from an initial point (source) to a final point (goal) in a weighted graph. To have a better comparison with the straight line shuttling case, we consider here that the start and end points of all shuttling paths have the same $y$ coordinates, $y(x=0)=y(x=300~\text{nm})$. The most important question is how the weight of each point in the $xy$ plane is defined. We consider two different ways of answering this question, one based on the SLZ model and the other based on the PLZ model. 

For the case based on the SLZ model, we consider that the weight at each point is given by $1/E_{\rm VS}$, which increases the cost of crossing regions with small valley splitting. The path obtained is shown by the green line in Fig.~\ref{2d}. The path deviates considerably from a straight line. Also, the entire range in the $y$ direction is used, which indicates that an even better path could be obtained by widening the 30~nm range in the $y$ direction. 

In the case based on the PLZ model, we define the weight at the $i$'th point on the $xy$ plane by the absolute value of the difference between the valley phases at the $i$'th point and the previous point in the path, $|\theta_i-\theta_{i-1}|$, which leads to a path that avoids valley phase changes. The optimized path in this case is shown as a cyan line in Fig.~\ref{2d}. This path deviates considerably from a straight line, and we have a much longer traveled distance compared to the straight-line case. 

\begin{figure}[h]
\includegraphics[width=\linewidth]{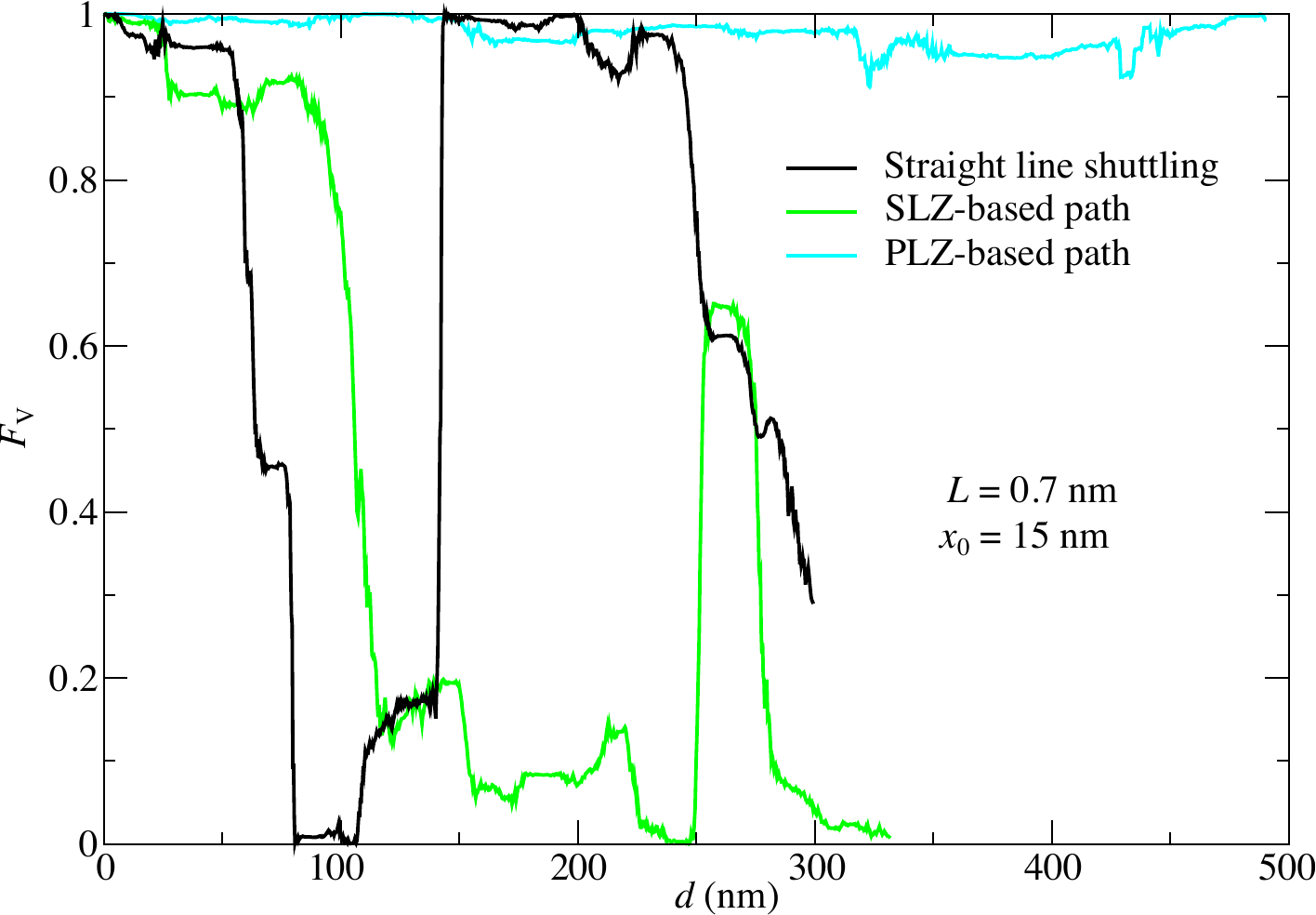}
\caption{\small The valley shuttling fidelity $F_V$ as a function of the travelled distance $d$ during shuttling for the device considered in Fig.~\ref{2d}. The black line represents  shuttling in a straight line, while the green and cyan lines correspond to the paths with the same colors in Fig.~\ref{2d}. The travelled distance during shuttling for a path that is not a straight line is longer than the 300~nm.} \label{fpath}
\end{figure}

The valley shuttling fidelity as a function of the shuttling distance for these different paths can be seen in Fig.~\ref{fpath}. Here, we are comparing the shuttling along a straight line  with the shuttling paths obtained by the lateral avoidance strategy based on the SLZ and PLZ models. In all cases, we consider a constant shuttling velocity of 5~mm/s. Although the lateral avoidance leads to a longer path compared to the straight line case, it suppresses the valley leakage during shuttling, with a much greater suppression obtained considering a strategy based on the PLZ model. This confirms, once more, that the partial Landau-Zener model describes the valley transitions during spin shuttling in Si quantum dots much better than the SLZ model.

Comparing with the previous strategy, we  conclude that an adapted velocity leads to a greater suppression of the valley leakage, which means that it is a better strategy than the lateral avoidance. The reason for this is that it is only possible to reduce rather than avoid the crossing of valley splitting minima during shuttling by deviating in the transversal direction, which explains the results based on the SLZ model. In the same way, the intervalley coupling still moves among the four quadrants of the complex plane during shuttling when the lateral avoidance strategy is used. This means that large values of $\dot{\theta}$ cannot be avoided here. For this reason, the lateral avoidance strategy is not as efficient as the adapted velocity. 

A different definition of the weight of each point in the $xy$ plane in the $A^*$ algorithm could lead to a better suppression of the valley leakage in this strategy. Furthermore, a combination of the strategies proposed here would also increase the valley shuttling fidelity. However, as already mentioned above, the DVL regime is not ideal for spin shuttling. Additionally, the realization of both strategies proposed here, the adapted velocity and the lateral avoidance, are a great experimental challenge. Therefore, even if these strategies could mitigate the valley leakage, the best strategy seems to be the spin shuttling in the FVT regime.

\section{Conclusions}
\label{sec:conclusions}

Valley excitations are considered one of the biggest challenges for a coherent spin shuttling in Si/SiGe heterostructures. In this work, we have investigated the transition dynamics between the two valley states during shuttling by considering a two-level system given by the two valley states during shuttling. We have identified that, depending on the shuttling velocity, the system can be in three distinct regimes. The distance-dependent valley leakage (DVL) regime is entered when shuttling slowly ($\lesssim 1$~m/s). In this regime, as the shuttling distance increases, the leakage between the valley states also increases. We have proposed strategies to suppress this leakage based on the partial Landau-Zener (PLZ) model, which proves to be much more efficient than previous strategies based on the standard Landau-Zener model. By increasing the shuttling velocity (approximately in the range of tens of m/s), we have found that the system changes to a new regime that we called full valley transitions (FVT) regime. Here, we have a sequence of full transitions between the two valley states during shuttling. In this case, there is no spin-valley entanglement and if a valley relaxation is avoided, inter-valley transitions would induce only a deterministic spin rotation that can be fixed after shuttling. The FVT regime is the ideal region in parameter space for deterministic spin shuttling, where it is possible to perform fast shuttling with no dephasing induced by valley excitation, independent of the shuttling distance. This result is in agreement with recent experimental work that obtained high spin shuttling fidelity after fast and long shuttling, revealing that valley excitations are not inducing the expected spin dephasing. By increasing the shuttling velocity even more (to hundreds of m/s and above) the system moves to the combined valley states (CVS) regime. Here, the electron will be in a linear combination of both valleys during shuttling, which induces spin dephasing that increases with the shuttling distance.

\begin{acknowledgments}
This work has been funded by the Federal Ministry of Education and Research (Germany), Grant No.~13N15657 (QUASAR),
and by Army Research Office (US) grant W911NF-15-1-0149.
\end{acknowledgments}

\end{document}